  \documentstyle[11pt,paspconf,psfig]{article}


\begin{document}

 \title{Microlensing, flattening,
$\Omega_B$ and $h$:
 from assumptions, via data to tentative conclusions }

\author{Vesna Milo\v sevi\'c-Zdjelar}

\affil{Dept. of Physics \& Astronomy, University of Manitoba, Winnipeg, Mb,  
R3T 2N2, Canada}

\author{Srdjan Samurovi\'c  and}

\affil{Dipartimento di Astronomia, Universit\`{a} degli Studi di Trieste,  Via Tiepolo 11, I-34131, Trieste, Italy}

\author{Milan M. \'Cirkovi\'c\altaffilmark{1}}
  
 \affil{Dept. of Physics \& Astronomy, SUNY at Stony Brook,
Stony Brook, NY 11794-3800, USA}



\altaffiltext{1}{Affiliated to: Astronomska Opservatorija, Volgina 7, 11000 Belgrade,
Serbia }

\begin{abstract}
 Microlensing observations can be used 
for determining the shape of the  Milky Way's halo.
It can be shown that the data are best described with 
moderately flattened halo,
$0.5{\:} ^< _\sim \:  q{\:}^< _\sim \: 0.6$. We 
discuss, by taking into account 
this result, the constraints on the baryonic mass-density
 parameter, $\Omega_B$,
and their implications on the value of the Hubble constant
$H_0$, i.e. $h$. 
Our conclusion is that current  data, in order to 
satisfy   BBNS and Galaxy dynamics constraints,
 strongly suggest that
$h$ should take values within the lower part of the
 permissible range, $h\sim 0.5$.

\end{abstract}

\section{Microlensing and the flattening of the Galactic halo}

We
 discuss  constraints on the  universal baryonic mass-density
 parameter $ \Omega_B \equiv {\rho_B / \rho_{\rm crit}}=8\pi G\rho _B/3H_0^2$ and a  Hubble parameter
$h \,\,\,  (H_0=100 \, h\; {\rm km/ s/ Mpc})$,  using
   all available microlensing (ML) data and  considering  flattened  
 Galactic dark halo. Since $\Omega _B$ and $h$ are uncertain and 
 correlated, we tried to break this degeneracy with the
 other cosmological
 information: measured optical depths from ML.
 In recent paper we examined the relation between
 galactic halo flattening and baryonic dark matter  (Samurovi\'c,
\'Cirkovi\'c \& Milo\v sevi\'c-Zdjelar 1999, 
S\'CMZ), using mass density for  flattened instead of spherical halo,  and  optical depth $\tau$  as  a function of
Galactic coordinates, derived by Sackett \&
Gould 1993  (eqs.  2.1 and 2.5). Comparing solutions for LMC, SMC, M31 and a 
 source  at 50 kpc with all available ML
data,  we obtained  $0.5{\:} ^< _\sim \:  q{\:}^< _\sim \: 0.6$ for halo flattening.  Assuming that MACHOs 
  have finite mass-to-light ratio (white or brown
dwarfs), in S\'CMZ we 
 determined   total mass of the MACHO halo as a function of flattening  (eq. 9),  and cosmological density parameter
$\Omega_{\rm MACHO}$ (eq. 10).

\section{Tentative conclusions}

Recent estimate of total halo mass  within 50 kpc is  
  $\sim  5.4^{+0.2}_{-3.6}\times 10^{11} M{}_\odot$
 (Wilkinson \& Evans 1999). That  can be 
  obtained when we assume
  flattening parameter $q\sim 0.6$. Spherical halo has serious problems concerning BBNS limits.

  Value  $h\sim 0.5$ is in accord
 with $h=0.39 ^{+0.14}_{-0.13}$ obtained by 
 combining the baryon budget with CMBR measurements
 (Hannestad 1999).  Realistic flattening and MACHO-dominance in the
inner halo  are incompatible with $ h > 0.75$.

\begin{figure}
 \begin{minipage}[t]{7cm}

\psfig{file=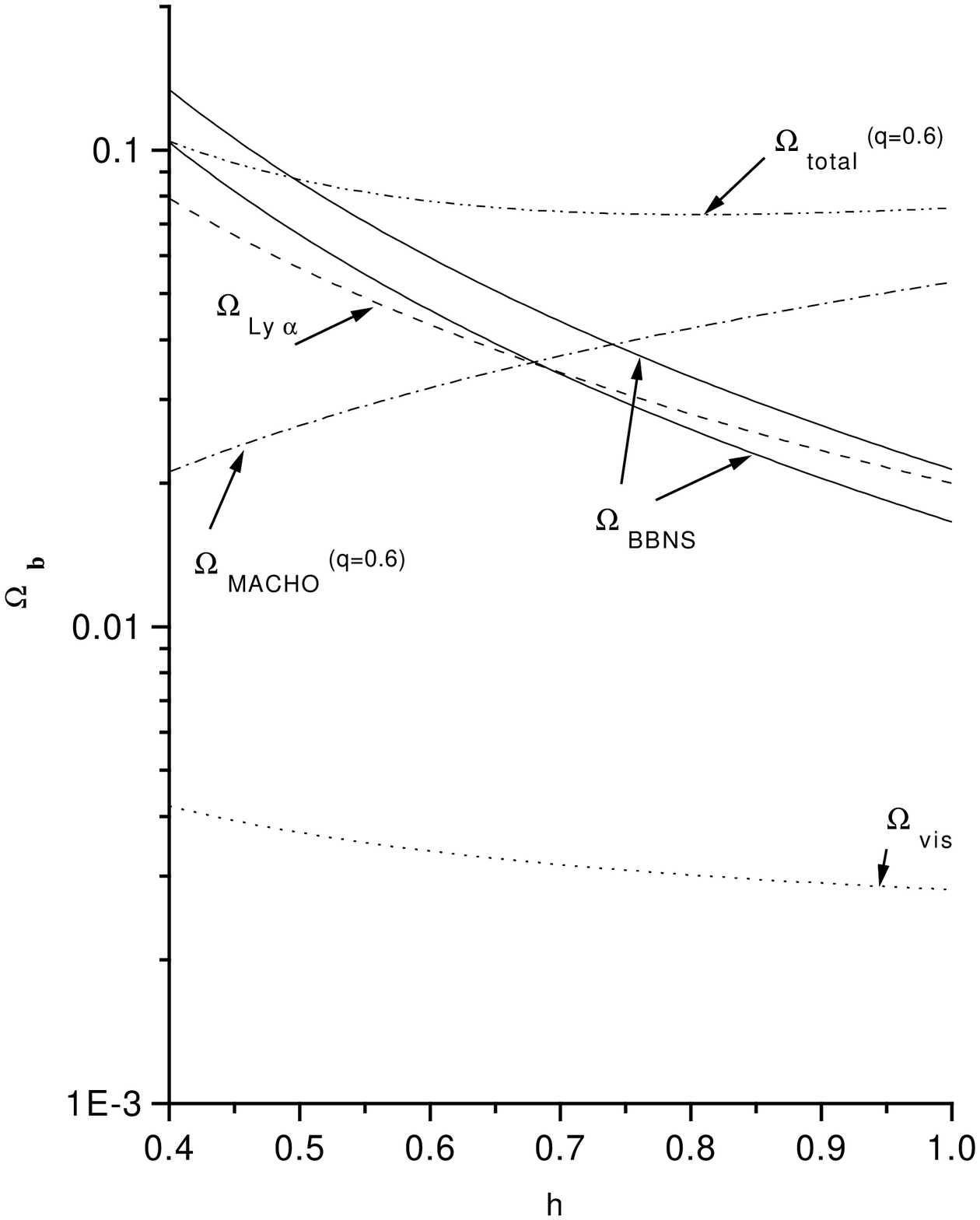,width=7cm,clip=}
\end{minipage}
\hfill
\begin{minipage}[t]{5.73cm}
\vspace*{-8.3cm}

{Fig. 1: Dependence of ${\Omega_B}$ on $h$ for flattening  
  $q=0.6$, truncation radius  50 kpc,  MACHO halo mass within 50 kpc for  $q=0.6$. The sum of main baryonic components is  $\Omega_{\rm total}$. High value 
 $0.09{\:} ^< _\sim \: {\Omega _B}{\:}^< _\sim \: 0.1$,  is  within  constraints  of  CMB observations: $\Omega _B < 0.28$ (Lineweaver et al. 1997), and recent BBNS and light element abundances constraints (Burles et al. 1999). At higher redshifts,  high $\Omega _B > 0.017  h^{-2}$ can account for observational results of Ly$\alpha$ forest. Our results  fit into theoretical requirements of high resolution spectroscopy of the IGM (e.g. Burles \& Tytler 1998). 
} 
\end{minipage}

\end{figure}

\acknowledgments
 Authors acknowledge financial support of the Univ. of Trieste, and kind encouragement of Dr.~Penny Sackett and Dr.~Geza Gyuk.

\end{document}